# Tunable cobalt vacancies and related properties in LaCo$_x$As$_2$

Shijie Shen,[†] Gang Wang,[*,†] Shifeng Jin,[†] Qingzhen Huang,[‡] Tianping Ying,[†] Dandan Li,[†] Xiaofang Lai,[†] Tingting Zhou,[†] Han Zhang,[†] Zhiping Lin,[†] Xiaozhi Wu,[†] Xiaolong Chen[*,†,§]

[†]Research & Development Center for Functional Crystals, Beijing National Laboratory for Condensed Matter Physics, Institute of Physics, Chinese Academy of Sciences, Beijing 100190, China
[‡]NIST Center for Neutron Research, NIST, Gaithersburg, MD 20899-6102, USA
[§]Collaborative Innovation Center of Quantum Matter, Beijing, China

**ABSTRACT:** The origin of transition metal vacancies and their effects on the properties of ThCr$_2$Si$_2$-type compounds have been less studied and poorly understood. Here we carefully investigate the structure, physical properties, and electronic structure for a series of lanthanum cobalt arsenides with nominal composition of LaCo$_x$As$_2$ (1.6 ≤ x ≤ 2.1). It is revealed that the occupancy of Co can be tuned between 1.98(1) and 1.61(1). The structural analyses based on x-ray and neutron diffractions show the existence of Co vacancies results from charge balance due to the formation of bond between As-As. These Co vacancies affect the magnetic and electrical properties greatly, adjusting the Curie temperature from 205 K to 47 K and increasing the resistivity by more than 100%. First-principles calculations indicate that the Co vacancies weaken the spin polarization and reduce the density of states at the Fermi level, resulting in decreased Curie temperature and increased resistivity, respectively. Our results address the importance of transition metal vacancies in ThCr$_2$Si$_2$-type materials and offer a reliable route to tune the magnetism of ThCr$_2$Si$_2$-type structure.

**Introduction**

The ThCr$_2$Si$_2$-type structure, composed of covalently bonded transition metal-metalloid layers and the intermediate metals, which are usually alkalis, alkali earths and lanthanides, is a common structure to around 1000 compounds.[1] Among them, 3d transition metal-contained compounds account for the majority. These compounds exhibit rich physical properties, such as the heavy fermion behavior,[2,3] the reentrant ferromagnetism,[4] the ferromagnetic quantum critical transition,[5] and superconductivity as well.[6-9] It is believed that the electron transfer from the intermediate metals to the layers and the valent state of 3d metals play important roles in determining the properties.

The electron transfer and the valent state of the 3d metal are highly correlated. If the amount of electron transfer exceeds what the layers can accommodate, with a 3d metal adopting its lowest stable valence, vacancies in the layers are expected. In this context, the formation of bonds between interlayer metalloids can also contribute to the occurrence of 3d metal vacancies in the layers. This is particular true for the late 3d metal-contained ThCr$_2$Si$_2$-type compounds. For instance, Ni, Cu, and Co vacancies are often observed in their corresponding compounds.[10-13] But the vacancies of Ni and Cu have no effect on the magnetism for Ni and Cu carry no localized magnetic moment. In contrast, the vacancies of Co can influence the magnetic property to some extent and have attracted interests. It has been reported that the existence of a little bit of Co vacancies lowers the Neel temperature from about 70 K to 52 K for CaCo$_2$As$_2$.[13] Recent studies

showed that Co vacancies exist in compounds RCo$_2$As$_2$ (R = La, Ce, Pr, Nd).[14,15]

Thompson *et al.*[14] demonstrated that LaCo$_2$As$_2$ undergoes a ferromagnetic transition at 86 K and highlighted the significant enhancement of $T_c$ (above 200 K), which is attributed to unintentional incorporation of a small amount of Bi and the formation of vacancies in the Co sublattice. The effect of incorporated Bi into the structure is emphasized.[14] However, it is not easy to understand that such a small amount of Bi (3.1%) could induce a significant increase in $T_c$. Here we report a careful investigation of crystal structure, physical properties, and electronic structure of LaCo$_x$As$_2$ as a function of Co occupancy for nominal 1.6 ≤ x ≤ 2.1. It is revealed that Co vacancies up to 19% can exist in collapsed tetragonal (cT) LaCo$_x$As$_2$ due to the formation of As-As bonding. We find that these Co vacancies in turn strengthen the As-As bonding, increase the resistivity, and most importantly, alter the temperature of magnetic ordering from 205 K to 47 K. The results show the possibility of tuning the magnetization of ThCr$_2$Si$_2$-type structure by transition metal vacancies.

**Experimental Section**

**Starting Materials.** Powder of As (Aladdin, 99.995%) was used as received. Finely dispersed chipping of La (Alfa Aesar, 99.9%) was used after carefully scraping away the surface layer. Co powder (Alfa Aesar, 99.8%) was additionally purified by heating under a flow of NH$_3$ gas at 500 °C for 30 min.

**Synthesis.** The starting materials were mixed in ratios of La: Co: As = 1.0: x: 2.0 (x = 2.1, 2.0, 1.9, 1.8, 1.7, and 1.6, total mass is about 1.5 g). Then they were pelletized and loaded in alumina crucibles. The crucibles were sealed inside silica tubes, which were backfilled with about 0.3 atm high purify argon gas. The samples were heated to 610 °C with a constant heating rate (100 °C/h), held there for 12 h (pre-reacting the volatile As with Co), then to 850 °C for 24 h (reacting the La chipping with As and Co to obtain a precursor for the second cycle). After carefully grinding, the samples were then heated at 1100 °C for 48 h and furnace-cooled to room temperature. All manipulations for sample preparation were carried out inside an argon-filled glove box (content of O$_2$ < 1 ppm).

**X-ray and Neutron Diffraction.** Room temperature powder x-ray diffraction (PXRD) data of nominal LaCo$_x$As$_2$ (x = 2.1, 2.0, 1.9, 1.8, 1.7, and 1.6) were collected using a PANalytical X'pert Pro diffractometer with Co Kα radiation(40 KV, 40 mA) and a graphite monochromator in a reflection mode (2$\vartheta$ = 10° to 130°, step = 0.017° (2$\vartheta$), and scan speed = 0.4 s/step). Neutron powder diffraction of nominal LaCo$_{1.8}$As$_2$ was performed (T = 300 K) at BT1 neutron powder diffractometer at NIST Center. A Cu (311) monochromator was used to produce the monochromatic neutron beam with the wavelength λ = 1.5398 Å. Collimations of 60', 15', and 7' were used before and after monochromator and after sample, respectively. Neutron powder diffraction of nominal LaCo$_{2.0}$As$_2$ was performed (T = 220 K and 4 K) with a Ge (733) monochromator with the wavelength λ = 1.1969 Å. Rietveld refinements of the data were performed with the FULLPROF package.[16] The structure of ThCr$_2$Si$_2$ was used as starting model for refinements of all these compounds. Only the *z* fraction coordinate of As atom is refinable among all atomic coordinates due to the restraint of the symmetry. During the refinements, a strong correlation between temperature factors and occupancy factors was observed, trying to refine the structures without restraint of temperature factors will lead to unreasonable results. So the temperature factors of La, Co, and As atoms were restraint to the values obtained from a single-crystal X-ray diffraction of La$_{0.97(1)}$Bi$_{0.03(1)}$Co$_{1.91(1)}$As$_2$[15] for all the refinements.

**Energy-Dispersive X-ray Spectroscopy (EDX).** The Co content of nominal LaCo$_x$As$_2$ (x = 2.1, 2.0, 1.9, 1.8, 1.7, and 1.6) samples was characterized by EDX. The result for each sample was obtained based on the average of three sets of data.

**Physical Property Measurements.** The static (DC) magnetic susceptibility of nominal LaCo$_x$As$_2$ (x = 2.1, 2.0, 1.9, 1.8, 1.7, and 1.6) was measured using a vibrating sample magnetometer (Quantum Design) in an applied field of 100 Oe at the temperatures ranging from 10 to 300 K. Temperature-dependent electrical resistivity measurements for nominal LaCo$_x$As$_2$ (x = 1.6 and 2.1), which were cold-pressed under a uniaxial stress of 400 kg cm$^{-2}$, were performed using the standard four-probe method on the physical property measurement system (Quantum Design) at the temperatures between 10 and 300 K.

**Electronic Structure Calculations.** Band structure calculations were performed using the CASTEP program code with plane-wave pseudopotential method.[17] We adopted the generalized gradient approximation with Perdew-Burke-Ernzerhof formula for the exchange-correlation potentials.[18] The ultrasoft pseudopotential with a plane-wave cutoff energy 330 eV and a Monkhorst Pack k-point separation of 0.04 Å$^{-1}$ in the reciprocal space were used for the calculations.[19] The self-consistent field was set as 5 × 10$^{-7}$ eV/atom. Full structural optimization including the lattice parameters and the atomic positions was performed with the convergence standard given as follows: energy change less than 5 × 10$^{-6}$ eV/atom, residual force less than 0.01 eV/Å, stress less than 0.02 GPa, and displacement of atom less than 5 × 10$^{-4}$ Å. Calculations were performed on two compounds, one is LaCo$_2$As$_2$ with complete CoAs layers, the other LaCo$_{1.5}$As$_2$ with ordered Co vacancies. Supercells √2 × √2 × 1 were built accordingly with two CoAs layers having total of eight Co/vacancy sites. Calculations on the magnetic order were performed assuming ferromagnetic ordering of all Co atoms within the structure.

**Results and Discussion**

Figure 1 displays the PXRD patterns using Co Kα radiation for a series of LaCo$_x$As$_2$ samples with nominal x of 2.1, 2.0, 1.9, 1.8, 1.7, and 1.6 collected at room temperature. The main phase can be well indexed on the body-centered tetragonal cell (space group *I*4/*mmm*). The side phase (marked by asterisk) is assigned to LaOCoAs, which is hard to be eliminated by various attempts. It is also noticed that side phase LaAs arises at nominal x = 1.6. As shown from the inset of Figure 1, the strongest peak of the main phase shifts gradually to higher angle with the decreasing nominal content of Co from 2.1 to 1.6. It manifests that the structure has a significant homogeneity range with respect to Co.

Figure 2 shows the Rietveld refinements for PXRD data of nominal LaCo$_{1.8}$As$_2$. The final agreement factors converge to $R_p$ = 2.45%, $R_{wp}$ = 3.20%, and $R_{exp}$ = 1.50%. Refinements of the site occupation factors result in significant deviation from unity for Co, but the La and As sites being fully occupied. The refined Co occupancy is 1.70(1). To explore the possibility of Co/As mix-occupancy, we did a concurrent refinement which takes into account the Co/As mix-occupancy on the Co-site and As-site and obtained both negative values, which manifest that Co/As mix-occupancy is unlikely in the compounds. Neutron diffraction (Figure S1a) confirms the existence of Co vacancies. The refined Co content of nominal LaCo$_{1.8}$As$_2$ is 1.72(3), which is consistent with the PXRD data. Rietveld refinements for other nominal LaCo$_x$As$_2$ (x = 2.1, 2.0, 1.9, 1.7, and 1.6) against PXRD data were also performed (Figure S2). All the crystallographic data including the molar ratios of phases are summarized in Table S1. What deserves to be mentioned is that the deficiency of Co is an intrinsic feature for LaCo$_x$As$_2$, the actual Co content is tunable between 1.98(1) and 1.61(1), with maximum vacancy ratio up to 19% (x = 1.61). EDX measurements were conducted to confirm the trend in variation of Co content across the series of samples. As shown in Table 1, the variation trend of Co content is in agreement with that of Rietveld refinements. The phenomenon of observed vacancies in these compounds is consistent with RCo$_2$As$_2$ (R = La, Ce, Pr, Nd), which were recently reported to have Co vacancies.[14,15]

The lattice constants and selected atomic distances for LaCo$_x$As$_2$ (x = 1.98, 1.92, 1.82, 1.70, 1.64, and 1.61) derived from structural refinements are schematized in Figure 3. Upon decreasing Co content, *a* axis slightly lengthens and *c* axis shortens by about 2%. The As atoms from adjacent layers get closer with As-As distance $d_{As-As}$ decreasing from 2.907(2) to 2.879(2) Å. The *c*/*a* ratio, obtained by the crystallographic data from Table S1, lies in the range from 2.60 to 2.53. These results indicate that LaCo$_x$As$_2$ (x = 1.98, 1.92, 1.82, 1.70, 1.64, and 1.61) are cT phases (which means that the interlayer anion-anion bonding pulls the layers closer and induces a relaxation of the in-plane lattice dimension.[20]) for their $d_{As-As}$ and *c*/*a* are very close to the respective values of the theoretically calculated and pressure-induced cT phase of CaFe$_2$As$_2$.[21-23] The strong contraction along *c* axis and expansion in *a* axis upon increasing vacancy levels can be interpreted from the charge balance point of view. Removal of Co reduces the net cationic charge in the material. This can be balanced by enhancing the As-As bonding ((As$_2$)$^{4-}$ has less formal negative charge than 2 × As$^{3-}$), which contracts *c*. In a similar manner, introduction of metal vacancy decreases the in-plane metal-metal bonding, which results in an expansion of *a* axis.

Now we would address the issue regarding the origin of metal vacancies in ThCr$_2$Si$_2$-type structure through a more comprehensive charge balance point of view. In ThCr$_2$Si$_2$-type structure, the transition metal-metalliod layer is negatively charged while the large-sized metal layer positively charged. A charge balance exists between them. In case the balance is tipped, charge compensation always takes into effect by two ways. One is the change in valance of the transition metals and the other is the introduction of vacancies either in the transition metal site or in the large-sized metal site. For example, when K is totally replaced by Sr in KCo$_2$As$_2$, a reduction in Co's valance is expected to maintain the charge balance. Moreover, when Sr is totally replaced by La in SrCo$_2$As$_2$, a further reduction in Co's valance should occur to form LaCo$_2$As$_2$ without Co vacancy like in LaCo$_2$P$_2$.[24-26] But since As-As bonding is present and Co is already in its observed lowest valance, the introduction of Co vacancy is the way to maintain the charge balance. For isostructural K$_x$Fe$_{2-y}$Se$_2$, vacancies are present both in K and Fe sites to maintain the charge balance.[28] This scenario also applies to compounds beyond the ThCr$_2$Si$_2$-type structure. A typical example is CeOCu$_{1-x}$Se where vacancies reside in Cu site.[31] This is due to the CeO layer provide more electrons than what CuSe layer can accommodate. The introduction of Cu vacancies is an effective way to maintain the charge balance. So the presence of vacancies is an effective means to keep the charge balance for a compound whose layer is in the state of carrier accommodation limit due to any factors that can induce charge changes. In our case, the As-As bonding is one of these factors responsible for the appearance of Co vacancies.

Figure 4 shows the temperature-dependent magnetic susceptibility of LaCo$_x$As$_2$ (x = 1.98, 1.92, 1.82, 1.70, 1.64, and 1.61). All of them clearly exhibit ferromagnetic ordering. The derivative of magnetic susceptibility shows that $T_c$ varies from 205 K to 47 K with the decreasing Co occupation from 1.98(1) to 1.61(1) (Figure S3). The side phase LaOCoAs with Curie temperature of about 60 K does not interfere with the ferromagnetic transition of LaCo$_x$As$_2$,[27,28] while LaAs is paramagnetic. The hysteresis loops at 10 K (Figure S4) indicate that all these compounds are soft ferromagnets with the coercive force ranging from about 2360 Oe to 50 Oe, which is similar to those for La$_{0.97}$Bi$_{0.03}$Co$_{1.9}$As$_2$ and LaCo$_2$P$_2$.[15,29] A substantial decrease of the saturation moment from 0.44 μ$_B$/Co to 0.24 μ$_B$/Co with increasing Co vacancies is observed from the isothermal field dependent magnetization. However, neutron diffraction data collected at 4 K fail to reveal the magnetic order of LaCo$_{1.92(1)}$As$_2$ (Figure S1b). As shown in Table 2, we compare the lattice constants and $T_c$ in the reference[14] to this work. It is found that the lattice constant and the Curie temperature of LaCo$_2$As$_2$ in reference[14] are close to the respective value of our sample LaCo$_{1.70(1)}$As$_2$. So the significantly enhanced $T_c$ of La$_{0.969}$Bi$_{0.031}$Co$_{1.91}$As$_2$ can be interpreted in a new light. It might be closely correlated with the lower Co deficiency.

The temperature-dependent resistivity for LaCo$_x$As$_2$ (x = 1.61 and 1.98) exhibits metallic behavior (Figure S5). LaCo$_{1.61(1)}$As$_2$ has a twice larger resistivity than that of LaCo$_{1.98(1)}$As$_2$. For LaCo$_{1.61(1)}$As$_2$, there is a clear quick decrease for resistivity at about 47 K, as shown in the inset of Figure S5, which is near $T_c$ of LaCo$_{1.61(1)}$As$_2$. However, the resistivity of LaCo$_{1.98(1)}$As$_2$ shows no anomaly near 200 K. This may be due to the fact that the resistivity contributed from electron-spin dispersion is very small in contrasts to the electron-lattice dispersion.

In order to understand the mechanism of decreased $T_c$ and increased resistivity with arising of Co vacancies, we turn to the first-principles calculations. The electronic structure of LaCo$_2$As$_2$ exhibits a high peak in the density of states (DOS) at the vicinity of the Fermi level (Figure S6), which accounts for the observed metallic behavior in Figure S5. It also shows that Co atoms contribute the majority DOS near the Fermi level. So the decline of Co content lowers the DOS at the Fermi level, resulting in the increase of resistivity. According to the Stoner criterion,[30] an itinerant magnet exhibits ferromagnetism when $I·N(E_F) > 1$, where $I$ is the strength of metal exchange interaction, which can be approximately obtained from the elemental metal,[31] $N(E_F)$ is the DOS at the Fermi level. The calculated value of $I·N(E_F)$ is 2.8 for LaCo$_2$As$_2$, which demonstrates the itinerant ferromagnetic ordering in this compound. This result is similar to the isostructural compound LaCo$_2$P$_2$.[29] Further calculations of spin-polarized DOS (Figure S7) show that the difference between the spin-up and spin-down DOS channels below the Fermi level is 0.97 states/cell for LaCo$_2$As$_2$. This value declines to 0.74 states/cell for LaCo$_{1.5}$As$_2$. This result means that intralayer Co-Co exchange interaction weakens with increasing Co vacancies, which is consistent with the observed elongated Co-Co distance and lowered $T_c$ in Co-deficient LaCo$_x$As$_2$.

**Conclusion**

In summary, we demonstrate that the non-stoichiometry is an intrinsic feature for LaCo$_2$As$_2$. The occupancy of Co can be tuned between 1.98(1) and 1.61(1) in LaCo$_x$As$_2$. We point out that the existence of Co vacancies results from charge balance due to the formation of bond between As-As. Co vacancies lower $T_c$ from 205 K to 47 K and increase the resistivity by more than 100%. The first-principles calculations indicate that LaCo$_2$As$_2$ is an itinerant ferromagnet. The occurrence of Co vacancies lowers the DOS at the Fermi level and weakens the spin polarization, which leads to the increased resistivity and decreased $T_c$, respectively. Knowledge of the key role of Co vacancies in the properties of LaCo$_x$As$_2$ allows the result of reference[14] to be interpreted in a new light. The results well address the importance of in-plane vacancies in the research area of superconductivity and related phenomena in ThCr$_2$Si$_2$-type materials and show the possibility of tuning the magnetism of ThCr$_2$Si$_2$-type structure through transition metal vacancies as well.

**AUTHOR INFORMATION**
**Corresponding Author**
gangwang@iphy.ac.cn
chenx29@iphy.ac.cn
**Notes**
The authors declare no competing financial interests.

**ACKNOWLEDGMENT**
S. J. Shen thanks Dr. J. Yan from Beihang University for helpful discussions about the neutron diffraction and Q. H. Zhang and X. Chen for assistance in the EDX measurements. This work was partly supported by the National Natural Science Foundation of China (Grant Nos. 51322211 and 90922037), the Strategic Priority Research Program (B) of the Chinese Academy of Sciences (Grant No. XDB07020100), Beijing Nova Program (Grant No. 2011096) and K. C. Wong Education Foundation, Hong Kong.

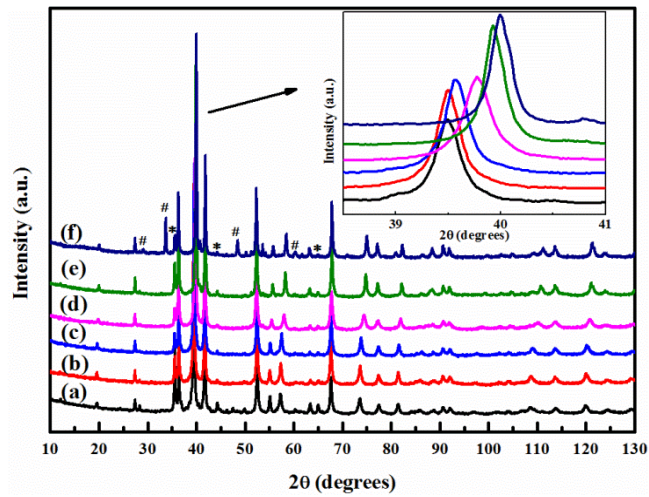

Figure 1. PXPD patterns of nominal LaCo$_x$As$_2$ (x = 2.1, 2.0, 1.9, 1.8, 1.7, and 1.6) (a-f), Co Kα radiation. The reflections marked by (*) and (#) are assigned to LaOCoAs and LaAs, respectively. The inset shows the enlarged (103) peak.

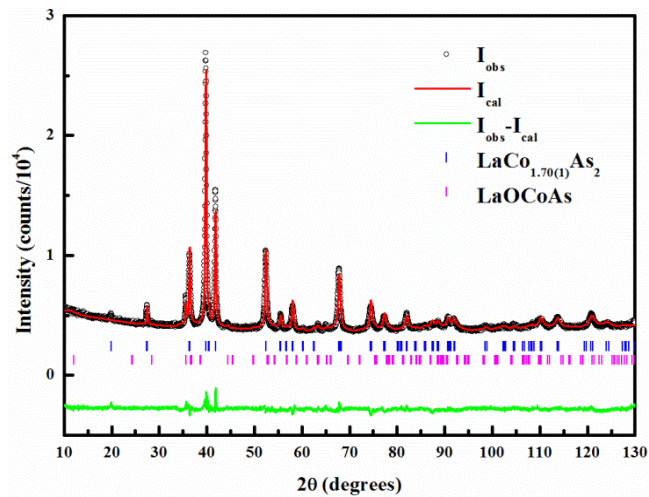

Figure 2. PXRD pattern of nominal LaCo$_{1.8}$As$_2$. The Rietveld refinement fits, difference profiles, and positions of Bragg peaks are also shown.

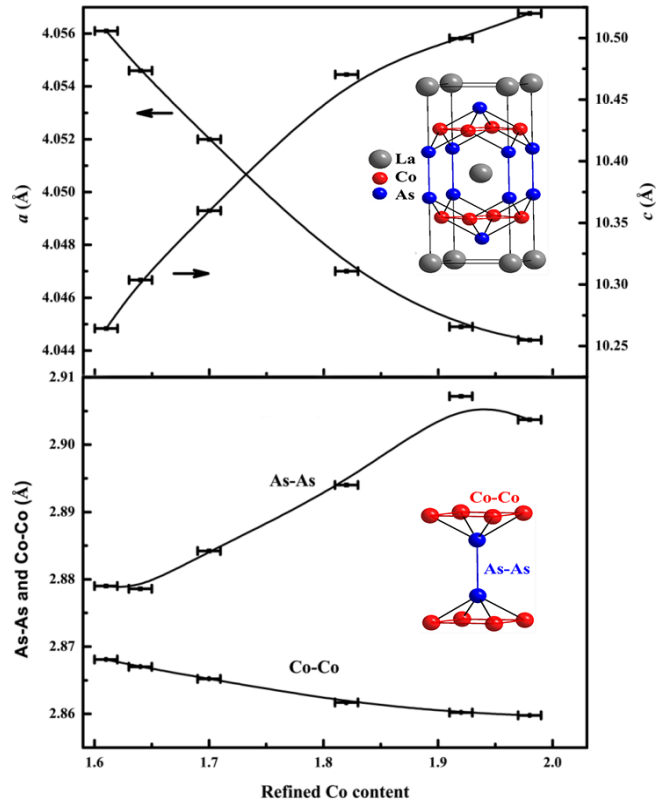

Figure 3. Lattice parameters, As-As and Co-Co distances of LaCo$_x$As$_2$ (x = 1.98, 1.92, 1.82, 1.70, 1.64, and 1.61) as a function of refined Co content.

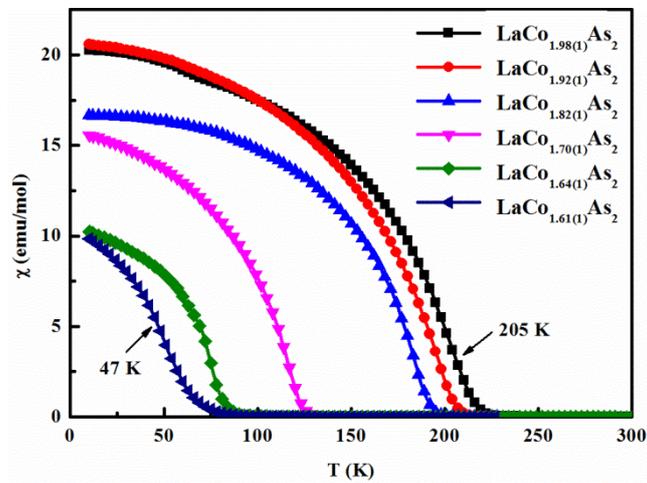

Figure 4. Temperature-dependent magnetic susceptibility of LaCo$_x$As$_2$ (x = 1.98, 1.92, 1.82, 1.70, 1.64, and 1.61) measured at 100 Oe. Arrows indicate the transition temperatures.

Table 1. Comparison of nominal, refined and EDX characterized Co content for nominal LaCo$_x$As$_2$ (x = 2.1, 2.0, 1.9, 1.8, 1.7, and 1.6) samples. The La content is set to 1.

| Nominal | Refined | EDX |
|---|---|---|
| **2.1** | 1.98(1) | 1.8(1) |
| **2.0** | 1.92(1) | 1.7(1) |
| **1.9** | 1.82(1) | 1.7(1) |
| **1.8** | 1.70(1) | 1.6(1) |
| **1.7** | 1.64(1) | 1.6(1) |
| **1.6** | 1.61(1) | 1.5(1) |

Table 2. Comparison of the lattice constants and $T_c$ between the reference and this work.

| Compound | $a$ (Å) | $c$ (Å) | $T_c$ (K) |
|---|---|---|---|
| **LaCo$_2$As$_2$** [a] | 4.0536(5) | 10.324(2) | 86 |
| **LaCo$_{1.70(1)}$As$_2$** [b] | 4.0520(3) | 10.3598(8) | 117 |
| **La$_{0.969}$Bi$_{0.031}$Co$_{1.91}$As$_2$** [a] | 4.0508(2) | 10.470(1) | 200 |
| **LaCo$_{1.98(1)}$As$_2$** [b] | 4.0444(3) | 10.5200 (8) | 205 |

[a] the data from the reference[14], [b] the data from this work.